\documentclass[prl,aps,amsfonts,amsmath,amssymb,twocolumn,superscriptaddress]{revtex4-2}

\usepackage{graphicx}
\usepackage{bm}
\usepackage{hyperref}
\usepackage{IEEEtrantools}
\usepackage{natbib}
\usepackage{float}
\usepackage{siunitx}
\usepackage{xcolor}

\restylefloat{table}

\begin{document}

\title{Observation of the massive Lee-Fukuyama phason in a charge density wave insulator}

\author{Soyeun Kim}
\affiliation{Department of Physics, University of Illinois at Urbana-Champaign, Urbana, IL 61801, USA}
\affiliation{Materials Research Laboratory, University of Illinois at Urbana-Champaign, Urbana, IL 61801, USA}

\author{Yinchuan Lv}
\affiliation{Department of Physics, University of Illinois at Urbana-Champaign, Urbana, IL 61801, USA}
\affiliation{Materials Research Laboratory, University of Illinois at Urbana-Champaign, Urbana, IL 61801, USA}

\author{Xiao-Qi Sun}
\affiliation{Department of Physics, University of Illinois at Urbana-Champaign, Urbana, IL 61801, USA}

\author{Chengxi Zhao}
\affiliation{Materials Research Laboratory, University of Illinois at Urbana-Champaign, Urbana, IL 61801, USA}\affiliation{Materials Science and Engineering Department, University of Illinois at Urbana-Champaign, Urbana, IL 61801, USA}

\author{Nina Bielinski}
\affiliation{Department of Physics, University of Illinois at Urbana-Champaign, Urbana, IL 61801, USA}
\affiliation{Materials Research Laboratory, University of Illinois at Urbana-Champaign, Urbana, IL 61801, USA}

\author{Azel Murzabekova}
\affiliation{Department of Physics, University of Illinois at Urbana-Champaign, Urbana, IL 61801, USA}
\affiliation{Materials Research Laboratory, University of Illinois at Urbana-Champaign, Urbana, IL 61801, USA}

\author{Kejian Qu}
\affiliation{Department of Physics, University of Illinois at Urbana-Champaign, Urbana, IL 61801, USA}
\affiliation{Materials Research Laboratory, University of Illinois at Urbana-Champaign, Urbana, IL 61801, USA}

\author{Ryan A. Duncan}
\affiliation{Stanford PULSE Institute, SLAC National Accelerator Laboratory, Menlo Park, CA 94025, United States}
\affiliation{Stanford Institute for Materials and Energy
Sciences, SLAC National Accelerator Laboratory, Menlo Park, CA 94025, USA}

\author{Quynh L. D. Nguyen}
\affiliation{Stanford PULSE Institute, SLAC National Accelerator Laboratory, Menlo Park, CA 94025, United States}
\affiliation{Stanford Institute for Materials and Energy
Sciences, SLAC National Accelerator Laboratory, Menlo Park, CA 94025, USA}

\author{Mariano Trigo}
\affiliation{Stanford PULSE Institute, SLAC National Accelerator Laboratory, Menlo Park, CA 94025, United States}
\affiliation{Stanford Institute for Materials and Energy
Sciences, SLAC National Accelerator Laboratory, Menlo Park, CA 94025, USA}

\author{Daniel P. Shoemaker}
\affiliation{Materials Research Laboratory, University of Illinois at Urbana-Champaign, Urbana, IL 61801, USA}
\affiliation{Materials Science and Engineering Department, University of Illinois at Urbana-Champaign, Urbana, IL 61801, USA}

\author{Barry Bradlyn}
\affiliation{Department of Physics, University of Illinois at Urbana-Champaign, Urbana, IL 61801, USA}

\author{Fahad Mahmood}
\email{fahad@illinois.edu}
\affiliation{Department of Physics, University of Illinois at Urbana-Champaign, Urbana, IL 61801, USA}
\affiliation{Materials Research Laboratory, University of Illinois at Urbana-Champaign, Urbana, IL 61801, USA}

\maketitle

\textbf{The lowest-lying fundamental excitation of an incommensurate charge density wave (CDW) material is
widely believed to be a \textit{massless} phason -- a collective modulation of the phase of the CDW order parameter. However, as first pointed out by Lee and Fukuyama \cite{Fukuyama1978}, long-range Coulomb interactions should push the phason energy up to the plasma energy of the CDW condensate, resulting in a \textit{massive} phason and a fully gapped spectrum. Whether such behavior occurs in a CDW system has been unresolved for more than four decades. Using time-domain THz emission spectroscopy, we investigate this issue in the material (TaSe$_4$)$_2$I, a classical example of a quasi-one-dimensional CDW insulator. Upon transient photoexcitation at low temperatures, we find the material strikingly emits coherent, narrow-band THz radiation. The frequency, polarization and temperature-dependence of the emitted radiation imply the existence of a phason that acquires mass by coupling to long-range Coulomb interaction. Our observations constitute the first direct evidence of the massive ``Lee-Fukuyama'' phason and highlight the potential applicability of fundamental collective modes of correlated materials as compact and robust sources of THz radiation.} 

\begin{figure*}[ht!]
\begin{center}
\includegraphics[width = 2\columnwidth]{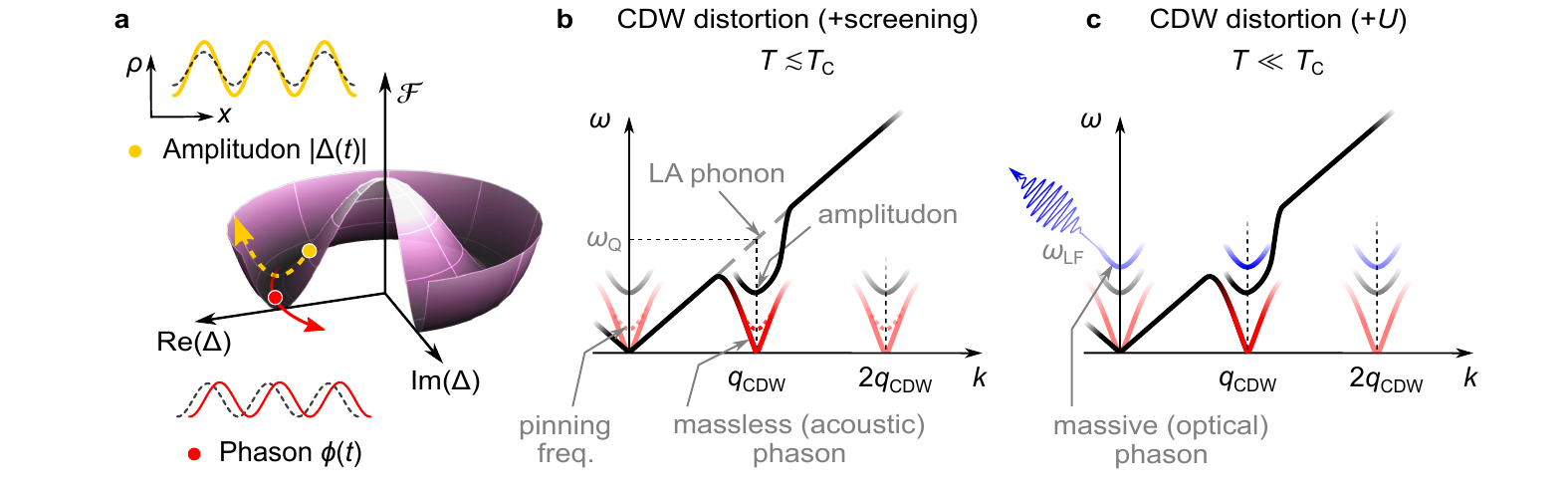}
\caption{Collective modes of an incommensurate charge-density-wave (CDW) phase. \textbf{(a)} Ginzburg-Landau free energy ($\mathcal{F}$) and real space representation of the amplitudon (dashed arrow) and the phason (solid line arrow) for a CDW order parameter. \textbf{(b-c)} Dispersion relations of the CDW collective modes. Below $T_\text{CDW}$, the otherwise undistorted acoustic phonon (LA) softens near $\vec{q}_\text{CDW}$ and renormalizes into the amplitudon and phason bands. \textbf{(b)} At moderate $T$ ($\lesssim T_\text{CDW}$) the long-range Coulomb repulsion $U$ is screened by quasiparticles. \textbf{(c)} When the system is cooled well below $T_\text{CDW}$, the screening weakens and the spectral weight from the massless (acoustic) phason is transferred to the massive (IR-active) phason of frequency $\omega_\text{LF}$.}
\label{fig-1}
\end{center}
\end{figure*}

The fundamental collective modes (amplitudon and phason) of a broken symmetry ordered state (Fig.~\ref{fig-1}a) have been key in establishing foundational theories across various fields of physics, including gauge theories in particle physics; and superconductors, antiferromagnets and charge-density wave (CDW) materials in condensed matter physics \cite{Schwartz2013, Fradkin2013, Coleman2015}. The phason is typically massless, in accordance with Goldstone's theorem, which necessitates the emergence of a massless boson for a broken symmetry in systems in which the ground or vacuum state is continuously degenerate. A prominent exception occurs in superconducting systems. Here, even though the ground state is continuously degenerate, the long-range Coulomb interaction pushes the longitudinal phason up to the plasma frequency~\cite{anderson1958coherent,anderson1958random}, and so a massless Goldstone boson does not exist and the low-lying excitation spectrum is fully gapped. This behavior is the celebrated Anderson-Higgs mechanism which established the deep connection between symmetry breaking and gauge fields, and ultimately explained how all fundamental particles acquire mass from interactions with the Higgs field.

Unlike superconductivity, an incommensurate CDW is believed to have a massless phason, typically understood in terms of softening of a longitudinal acoustic phonon branch around the CDW wavevector $\vec{q}_\text{CDW}$ (Fig.~\ref{fig-1}b). Below the CDW transition temperature ($T_\text{CDW}$), this mode softening results in the linear-in-wavevector, zero-gap dispersion of the phason, implying that the CDW can freely slide for excitation wavevector $\vec{q} = 0$. In any real material, however, random impurities and disorder restrict this sliding motion, leading to a small gap in the phason dispersion (the pinning frequency) (Fig.~\ref{fig-1}b). Thus, the sliding CDW motion can only be observed if a strong enough electric field is applied to first depin the phason. The resulting sliding motion of the CDW can then be measured in DC transport experiments as has been done in various systems \cite{Gruner1988_review}.

However, the above phenomenology of a massless phason (or disorder pinned phason at low frequency) assumes the absence of long-range Coulomb interactions ($U$). This assumption is believed to be valid because the presence of normal electrons at a non-zero temperature can screen $U$. However, if $U$ were sufficiently strong, or if the density of normal electrons were sufficiently low, then the CDW phason at $\vec{q} = 0$ should be pushed to higher energies (even above the amplitudon energy) (Fig.~\ref{fig-1}c). This behavior was highlighted in the seminal work of Lee, Rice and Anderson (LRA) on CDW dynamics \cite{LRA1973,LRA1993s1} and soon formalized by Lee and Fukuyama~\cite{Fukuyama1978} who noted the similarity of the emergence of the massive CDW phason with the Anderson-Higgs mechanism in a superconductor. Later works~\cite{Takada1987,Maki1993} predicted that the massive (optical) phason could indeed dominate over the massless (acoustic) phason at sufficiently low temperatures where charged quasiparticles cannot sufficiently screen $U$ \cite{Takada1987,Maki1993}. We note that in superconductors the plasma frequency is much larger than the single particle gap, rendering the phase mode unobservable deep in the superconducting phase. In a CDW system, however, the relevant scale is the plasma frequency of the condensate, which can lie far below the single-particle gap. To date, direct experimental evidence of the massive ``Lee-Fukuyama'' phason in CDW systems has been elusive.

\begin{figure*}[ht!]
\begin{center}
\includegraphics[width = 2\columnwidth]{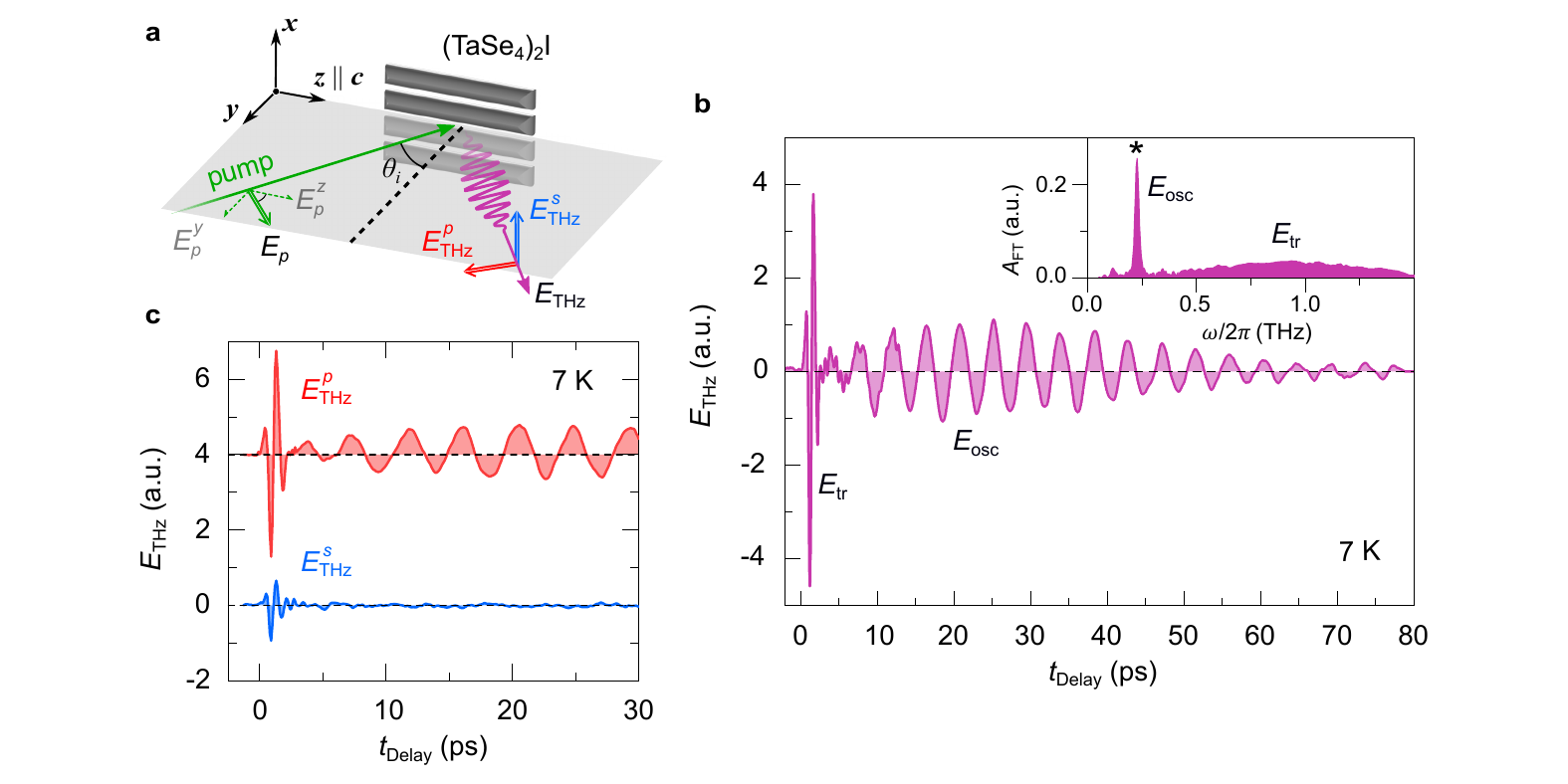}
\caption{Terahertz (THz) emission from (TaSe$_4$)$_2$I. \textbf{(a)} The geometry of the sample orientation, incident light (pump) and emitted THz polarizations. The TaSe$_4$ chains ($c$-axis) are oriented along the $z$-axis, and the pump beam is ${p}$-polarized with incident angle $\theta_i = 45^\circ$. The plane of incidence is represented as the gray shaded region. \textbf{(b)} The THz emission signal ($E_\text{THz}$) as a function of time ($t_\text{delay}$), measured at 7 K. The Fourier transform amplitude ($A_\text{FT}$) is plotted in the inset. The 0.23 THz mode is marked with an asterisk (*). \textbf{(c)} The $p$- and $s$-polarized components of the THz emission signal, $E_\text{THz}^p$ (upper) and $E_\text{THz}^s$ (lower), respectively. $E_\text{THz}^p$ is shown with an offset.}
\label{fig-2}
\end{center}
\end{figure*}

Here, we present evidence for the generation and detection of a massive Lee-Fukuyama phason in the CDW insulator (TaSe$_4$)$_2$I, a quasi-1D material \cite{Gressier1982} that undergoes an incommensurate CDW transition below $T_\text{CDW} \sim $ 260 K \cite{Wang1983,Maki1983,Fujishita1984,Lee1985} with a CDW gap for single particle excitations of 2$\Delta_\text{CDW}\approx 250-300$meV \cite{Donovan1990, Degiorgi1991, Gooth2019,Soyeun2021}. (TaSe$_4$)$_2$I is unique among quasi-1D CDW systems due to its unusually high resistivity in the low temperature insulating state \cite{Suppl}, such that the long-range Coulomb interaction can remain unscreened and the Lee-Fukuyama phason can have have significant spectral weight. To investigate the collective modes of the CDW order in (TaSe$_4$)$_2$I, we performed time-domain THz emission spectroscopy using an ultrafast photoexcitation `pump' pulse with an energy of 1.2 eV ($\lambda$ = 1030 nm) (See Fig.~\ref{fig-2}a for the experimental geometry and \cite{Suppl} for further details of the experimental setup). Note that the photoexcitation energy here is greater than 2$\Delta_\text{CDW}$ and so the pump pulse initially creates single-particle excitations across the CDW gap. As (TaSe$_4$)$_2$I is structurally chiral (SG. 97) and lacks inversion symmetry, this photoexciation generates a transient current with a duration of a few picoseconds (ps). Such phenomena is well-known as a photo-galvanic or a photo-Dember effect, both of which can occur in systems lacking inversion symmetry \cite{Belinicher1980, Sakai2005}. The transient current then results in a short ps-duration burst of THz radiation in the far-field which we measure in the time-domain using standard electro-optical sampling (EOS) \cite{Suppl}. 

Figure~\ref{fig-2}b shows the measured time profile of the THz electric field (${E}_\text{THz}(t)$) emitted from (TaSe$_4$)$_2$I upon photoexcitation at $T$ = 7 K $\ll$ $T_\text{CDW}$. Here the pump is $p$-polarized, with an electric field component along the quasi-1D chains of (TaSe$_4$)$_2$I. Two features are immediately evident in $E_\text{THz}(t)$: a transient peak around $t_\text{delay}=0$ ps followed by a long-lived coherent oscillation that lasts over $\sim$ 80 ps. In the frequency domain (Fig.~\ref{fig-2}b inset), the transient peak around $t_\text{delay} \sim 0$ ps manifests as a broad distribution over frequencies from 0.1 to 2 THz, while the long-lived coherent oscillation manifests as a sharp peak centered at 0.23 THz. For the rest of this work, we refer to the transient ($t_\text{delay} \sim 0$ ps) peak as $E_\text{tr}$ and the coherent oscillation as $E_\text{osc}$. As noted above, $E_\text{tr}$ is what we typically expect from such an experiment due to a photo-galvanic or a photo-Dember effect. The transient current can be estimated from the measured $E_\text{tr}$, and is greater than the current necessary to depin the dynamics of CDW order in (TaSe$_4$)$_2$I \cite{Suppl}.

\begin{figure*}[ht!]
\begin{center}
\includegraphics[width = 1.95\columnwidth]{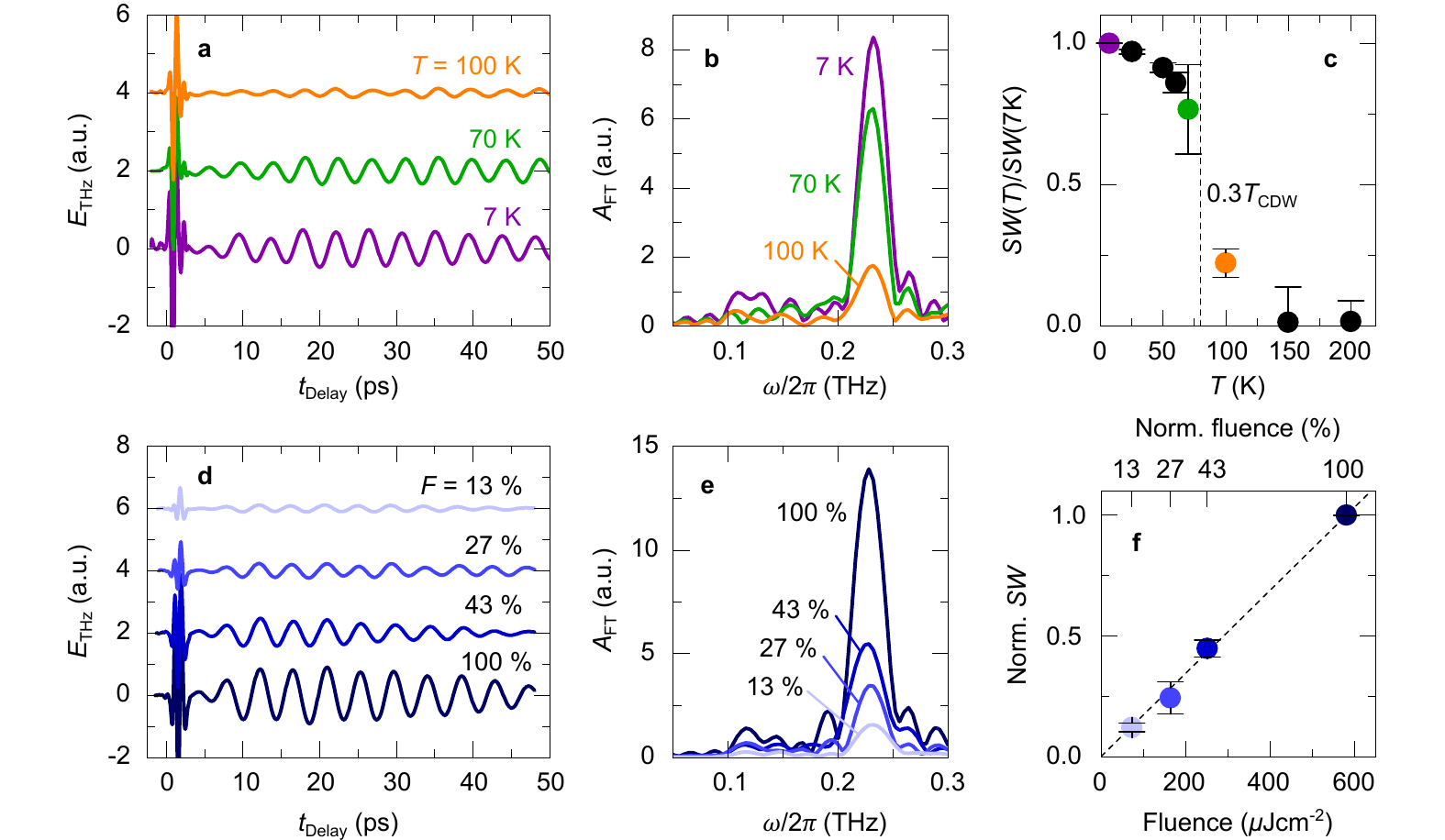}
\caption{Temperature ($T$) and pump fluence ($F$) dependence of the 0.23 THz mode. \textbf{(a)} THz emission signal as a function of delay time at a few selected temperatures. The signals at different temperatures are offset for clarity. \textbf{(b)} Fourier transforms of the traces shown in \textbf{(a)}. \textbf{(c)} Spectral weight ($SW$) of the 0.23 THz mode as a function of temperature, normalized to the $SW$ at 7 K. Dashed line indicates 0.3$T_\text{CDW} \sim$ 80 K. \textbf{(d)} THz emission signal at 7 K as a function of delay time at a few selected pump fluences. The signals at different fluences are shown with offsets for clarity. \textbf{(e)} Fourier transforms for the traces shown in \textbf{(d)}. \textbf{(f)} $SW$ of the 0.23 THz mode as a function of pump fluence, normalized to the maximum value of the $SW$.}
\label{fig-3}
\end{center}
\end{figure*}
 
We focus on the observed narrow-band THz emission $E_\text{osc}$ since this aspect of the data is particularly striking. $E_\text{osc}$ lasts well over 80 ps -- much longer than the typical few ps duration signal expected from semiconductors \cite{Sakai2005} and semimetals \cite{Rees2020, Gao2020, Luo2021} in THz emission experiments. Another unusual feature of $E_\text{osc}$ is the observed waveform envelope in the time domain which appears to gradually increase in magnitude starting at $t_\text{delay}$ = 0 ps. In impulsive excitation ultrafast experiments, the signal is usually peaked at $t_\text{delay}$ = 0 from where it exponentially decays. Additionally, while the measured transient peak $E_\text{tr}$ is both horizontally and vertically polarized, the coherent oscillation $E_\text{osc}$ is only horizontally polarized (Fig.~\ref{fig-2}c). In our experimental geometry, this corresponds to $E_\text{osc}$ being polarized along the quasi-1D chains of (TaSe$_4$)$_2$I as shown in Fig.~\ref{fig-2}a.

To investigate the origin of the radiating mode, we study the $E_\text{osc}$ spectra at various different temperatures (Figs.~\ref{fig-3}a,b). The $E_\text{osc}$ signal is strongest at 7 K and gradually decreases with increasing temperature, showing a sudden drop at $\sim$ 80 K (at about 30\% of $T_\text{CDW}$). This feature is also clear in the integrated spectral weight of the 0.23 THz mode as a function of temperature shown in Fig.~\ref{fig-3}c. Note that the initial transient signal, $E_\text{tr}$, is similar at both 70 K and 100 K but the oscillating mode $E_\text{osc}$ is significantly weaker at 100 K. To further understand the origin of the radiating mode, we studied the dependence of $E_\text{THz}$ on the incident pump fluence, shown in Figs.~\ref{fig-3}d-f. The spectral shape of $E_\text{osc}$ does not change with decreasing pump fluence - only the overall spectral weight decreases linearly with decreasing fluence (Fig.~\ref{fig-3}f), indicating that we are in a perturbative regime.

Based on the observations above, we can rule out several possible sources for the radiating mode at 0.23 THz, such as optical phonons~\cite{Dekorsy1996,Sakai2005}, longitudinal acoustic phonons~\cite{Schaefer2013} or the CDW amplitudon. The lowest-frequency optical phonons in (TaSe$_4$)$_2$I are at much higher frequencies (1.1 THz \cite{Lorenzo1993}) and THz oscillatory emission from optical or acoustic phonons would be peaked at $t_\text{delay} \sim 0$ ps followed by an exponential decay, unlike the waveform shape we observe. The amplitudon in (TaSe$_4$)$_2$I cannot have a net IR-dipole moment (due to the $D_2$ (222) point group symmetry of the incommensurate phase~\cite{Shi2021}) which is necessary for the far-field coherent radiation observed here. Furthermore, the spectral weight of THz emission due to phonons or the amplitude mode is not expected to decrease to zero near $T\sim$ 80 K. We also note, that the relatively low pump fluences used in our experiments ($\sim$ 580 $\mu$Jcm$^{-2}$) and the linear pump fluence dependence precludes explanations in terms of a complete melting and recovery of CDW order, which would have resulted in a saturating-type behavior with pump fluence.

\begin{figure}[hbt!]
\begin{center}
\includegraphics[width = 0.9\columnwidth]{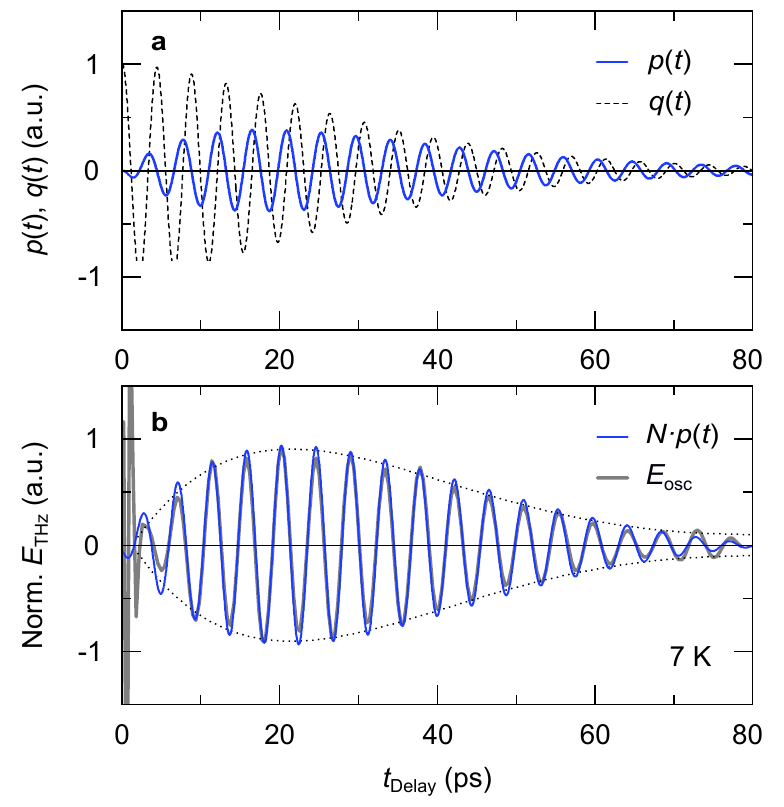}
\caption{Model of a radiating massive Lee-Fukuyama phason coupled to a non-radiating acoustic phonon. \textbf{(a)} $p(t)$ and $q(t)$
represent the time-profile of the massive phason and acoustic phonon, respectively, as obtained from a solution of a coupled harmonic oscillators model~\cite{Suppl}. This particular set corresponds to $p(t)$ obtained from a fit to the data at 7 K as shown in \textbf{(b)}. $N$ is a constant to rescale $p(t)$.}
\label{fig-4}
\end{center}
\end{figure}

Having ruled out the possible sources discussed above, we posit that the massive Lee-Fukuyama phason in (TaSe$_4$)$_2$I gives the coherent radiation at 0.23 THz. Within the LRA model for CDW dynamics \cite{LRA1973, Fukuyama1978}, the massive Lee-Fukuyama phason frequency $\omega_\text{LF}$ at $\vec{q}= 0$ is given as $\omega_\text{LF}^2 = \frac{3}{2}\lambda\omega_\text{Q}^2$, where $\omega_\text{Q}$ is the frequency of the normal state longitudinal acoustic phonon at $\vec{q}_\text{CDW}$ (Fig.~\ref{fig-1}b) and $\lambda$ is the electron-phonon coupling constant. For (TaSe$_4$)$_2$I, $\omega_\text{Q} \approx 0.25$ THz, as measured using inelastic neutron scattering \cite{Fujishita1986, Lorenzo1998} while $\lambda \sim 0.6$ \cite{Kim1991} for the strong electron-phonon coupling in (TaSe$_4$)$_2$I. This gives an expected $\omega_\text{LF} \approx 0.24$ THz, close to our observed mode frequency of 0.23 THz. Additionally, the temperature dependence of the 0.23 THz mode, which shows a considerable spectral weight only for $T \lesssim$ 0.3 $T_\text{CDW}$, agrees quite well with the random phase approximation (RPA) calculations of Virosztek and Maki \cite{Maki1993} for the temperature dependence of a massive phason. Taken together, our observations strongly suggest that the narrow-band 0.23 THz radiation originates from the predicted massive Lee-Fukuyama phason in a CDW insulator. 

To determine the excitation mechanism for the massive phason, we further discuss the temporal waveform of the $E_\text{osc}$ highlighted earlier. Note that the $E_\text{osc}$ shown in Fig.~\ref{fig-2}b is roughly zero at $t_\text{delay}$ = 0 ps and then becomes significant only at later times. We fit the behavior of $E_\text{osc}$ in time using a model of two coupled modes, one of which is excited at $t_\text{delay}$ = 0 ps while only the other one radiates as in Fig.~\ref{fig-4} (See \cite{Suppl} for details on the fitting procedure). This fitting model is consistent with a picture where the radiating mode is the $\vec{q}=0$ massive phason with frequency $\omega_{\mathrm{LF}}$, while the non-radiating mode is an acoustic phonon mode with wavevector $\vec{q}=\vec{q}_\text{CDW}$ and frequency $\omega_\mathrm{Q}\approx \omega_{\mathrm{LF}}$. The presence of such an acoustic mode in the CDW phase was established in Refs.~\cite{Lorenzo1998, Schaefer2013}. Since the acoustic phonon frequency $\omega_{\mathrm{Q}} \ll |\vec{q}_\text{CDW}|c$ is well outside the light cone for free space radiation, the acoustic phonon cannot directly radiate. However, since it has wavevector $\vec{q}=\vec{q}_\text{CDW}$, it can scatter off the CDW modulation and excite the massive phason mode. The massive phason mode at $\vec{q}=0$ will then radiate, explaining the time profile of $\vec{E}_\text{osc}$ in Figs.~\ref{fig-3} and \ref{fig-4}. We hypothesize that the incident IR pump excites a large number of single-particle excitations above the CDW gap. These excitations rapidly relax via the generation of high-energy phonons which in turn eventually decay into acoustic phonons. The low-lying acoustic phonon with $\vec{q}=\vec{q}_\text{CDW}$ and $\omega_\mathrm{Q}\approx\omega_{\mathrm{LF}}$ will then slowly decay via scattering into the massive phason.

We note that (TaSe$_{4}$)$_2$I was recently argued to be an axion insulator (in the presence of an external magnetic field) based on magnetoconductance measurements of the sliding phason dynamics \cite{Gooth2019}, though such an interpretation is under active debate \cite{Sinchenko2022}. Since (TaSe$_4$)$_2$I is a good insulator at low temperatures \cite{Suppl}, the reported magnetoconductance measurements \cite{Gooth2019} were limited to temperatures above 80 K. Our results imply that at low temperature long-range Couloumb interaction is crucial for a full understanding of the CDW order in (TaSe$_4$)$_2$I and related axion insulator candidates. Contributions to the longitudinal magnetoconductance due to axion electrodynamics may occur near the massive phason frequency, which can provide a promising way to manipulate axionic states coherently.

To conclude, we emphasize that converting an ultrafast infrared pulse into a multicycle THz pulse with a narrow bandwidth of a few GHz, as demonstrated here, is in sharp contrast to existing ultrafast-based emitters that produce single-cycle, broadband THz pulses. Our work thus reveals the possibility of using fundamental collective modes as sources of robust and narrow-band THz radiation. Such sources would enable resonant control of light-matter interaction and narrow-band spectroscopy at the intrinsic energy scales of correlated materials. Furthermore, our methodology here can provide a direct dynamical probe of collective excitations and effects of long-range interactions in materials with modulated order parameters such as spin or pair density waves.

\vspace{2mm}

\noindent\textbf{Acknowledgements:} We thank Peter Abbamonte, Peter Armitage, Dipanjan Chaudhuri, Tai-Chang Chiang, Lance Cooper, Steve Kivelson, Anshul Kogar, Patrick Lee, Vidya Madhavan, Darius Torchinsky, and Benjamin Wieder for insightful discussions.\\
This work was supported by the Quantum Sensing and Quantum Materials, an Energy Frontier Research Center funded by the U.S. Department of Energy (DOE), Office of Science, Basic Energy Sciences (BES), under Award No.DE-SC0021238. F.M. acknowledges support from the EPiQS program of the Gordon and Betty Moore Foundation, Grant GBMF11069. R.A.D. acknowledges support from the Bloch Postdoctoral Fellowship in Quantum Science and Engineering of the Stanford University Quantum Fundamentals, Architectures, and Machines initiative (Q-FARM), and from the Karel Urbanek and Marvin Chodorow Postdoctoral Fellowship of the Stanford University Department of Applied Physics. The authors acknowledge the use of the spectroscopic ellipsometry setup at the Institute for Basic Science (IBS) in Korea (Grant No. IBS-R009-D1).\\

\vspace{2mm}

\noindent\textbf{Data availability.} All relevant data are available on reasonable request.

\vspace{2mm}

\noindent\textbf{Competing financial interests:} The authors declare no competing financial interests.

\vspace{2mm}

\noindent\textbf{Author contributions:} S.K., Y.L., N.B., A.M. and F.M. performed the THz emission spectroscopy experiments and the corresponding data analysis. X-Q.S and B.B. developed the theoretical understanding and modelling. C.Z., K.Q. and D.P.S synthesized and characterized the samples. R.A.D., Q.L.D.N., and M.T. gave crucial insights to the understanding and analysis of the data. S.K., X-Q.S, B.B. and F.M. wrote the manuscript with input from all the authors. F.M. conceived and supervised this project.

\bibliographystyle{naturemag}
\bibliography{References.bib}
\end{document}